\documentclass[aps,prstab,preprint,groupedaddress]{revtex4}
\usepackage{graphicx}

\begin{document}


\preprint{SLAC-AP-141}
\preprint{May 2002}


\title{An Accurate, Simplified Model of Intrabeam Scattering}


\author{Karl L.F. Bane}
\thanks{Work supported by the Department
of Energy, contract DE-AC03-76SF00515}
\affiliation{Stanford Linear Accelerator Center,\\
Stanford University, Stanford, CA 94309}



\begin{abstract}

Beginning with the general Bjorken-Mtingwa solution for intrabeam
scattering (IBS) we derive an accurate, greatly simplified model
of IBS, valid
for high energy beams in normal storage ring lattices. In addition, we
show that, under the same conditions, a modified version of Piwinski's
IBS formulation (where $\eta^2_{x,y}/\beta_{x,y}$ has been
replaced by ${\cal H}_{x,y}$) asymptotically approaches the
result of Bjorken-Mtingwa.

\end{abstract}


\maketitle

\section*{INTRODUCTION}

Intrabeam scattering (IBS), an effect that tends to increase the beam emittance,
is important
in hadronic\cite{Bhat:99} and heavy ion\cite{RHIC:01} circular machines, as well
as in low emittance electron storage rings\cite{ATF:02b}. In the former
type of machines it results in emittances that continually
increase with time; in the latter type, in steady-state
emittances that are larger than those
given by quantum excitation/synchrotron radiation
alone.

The theory of intrabeam scattering for accelerators was first developed by
Piwinski\cite{Piwinski:74}, a result that was extended by
Martini\cite{Martini:84}, to give a formulation that we call here
the standard Piwinski~(P) method\cite{Piwinski:99}; this was
followed by the equally detailed Bjorken and Mtingwa (B-M)
result\cite{Bjorken:83}. Both approaches solve the local,
two-particle Coulomb scattering problem for (six-dimensional)
Gaussian, uncoupled beams, but the two results appear to be
different; of the two, the B-M result is thought to be the more
general\cite{Piwinski:p}.

For both the P and the B-M methods solving
for the IBS growth rates is time consuming,
involving, at each time (or iteration) step, a numerical integration
at every lattice element.
Therefore, simpler, more approximate
formulations of IBS have been developed over the years: there are
approximate solutions of
Parzen\cite{Parzen:87}, Le Duff\cite{LeDuff:89},
Raubenheimer\cite{Raubenheimer:91}, and Wei\cite{Wei:93}.
In the present report we
derive---starting with the general B-M formalism---another
approximation, one accurate and valid for
high energy beams in normal storage ring lattices.
We, in addition, demonstrate that under these same conditions
a modified version of Piwinski's IBS formulation asymptotically becomes equal to
this result.


\section*{HIGH ENERGY APPROXIMATION TO BJORKEN-MTINGWA}

\subsection*{The General B-M Solution\cite{Bjorken:83}}

Let us consider first machines with bunched beams that are uncoupled
and have vertical dispersion due to {\it e.g.} orbit errors.
Let the intrabeam scattering growth rates be defined as
\begin{equation}
{1\over T_p}={1\over\sigma_p}{d\sigma_p\over dt}\ ,\quad {1\over
T_x}={1\over\epsilon_x^{1/2}}{d\epsilon_x^{1/2}\over dt}\ ,\quad
{1\over T_y}={1\over\epsilon_y^{1/2}}{d\epsilon_y^{1/2}\over dt}\ ,
\label{growth_def_eq}
\end{equation}
with $\sigma_p$ the relative energy spread, $\epsilon_x$ the
horizontal emittance, and $\epsilon_y$ the vertical emittance.
The growth rates according to Bjorken-Mtingwa
(including a $\sqrt{2}$ correction factor\cite{Kubo:01b}, and including
vertical dispersion) are
\begin{eqnarray}
&&\hspace{-16pt}{1\over T_i}\ =\ 4\pi A({\rm log})\bigg<\int_0^\infty {d\lambda\,
\lambda^{1/2}\over [{\rm det}(L+\lambda I)]^{1/2}}\bigg\{
\nonumber\\
&&\hspace{-8pt}
 TrL^{(i)}Tr\left({1\over L+\lambda I}\right)
 -\ 3TrL^{(i)}
\left({1\over L+\lambda I}\right)\bigg\}\bigg>\quad\ \
\label{BM_eq}
\end{eqnarray}
where $i$ represents $p$, $x$, or $y$;
\begin{equation}
A= {r_0^2c N\over 64\pi^2\bar\beta^3\gamma^4\epsilon_x\epsilon_y\sigma_s\sigma_p}
\quad,
\end{equation}
with $r_0=2.82\times10^{-15}$~m, the classical electron radius,
$c$ the speed of light, $N$ the bunch population,
$\bar\beta$ the velocity over $c$, $\gamma$ the Lorentz energy factor,
and $\sigma_s$ the bunch length;
$({\rm log})$ represents the Coulomb log factor,
$\langle\rangle$ means that the enclosed
quantities, combinations of beam parameters and lattice
properties,
 are averaged around the entire ring; ${\rm det}$ and $Tr$ signify, respectively, the determinant and the
trace of a matrix, and $I$ is the unit matrix.
Auxiliary matrices are defined as
\begin{equation}
L=  L^{(p)} + L^{(x)} + L^{(y)} \quad,
\end{equation}
\begin{equation}
L^{(p)}={\gamma^2\over\sigma_p^2}\left(
\begin{array}{ccc}
0 & 0 & 0 \\
0 & 1 & 0 \\
0 & 0 & 0
\end{array}\right)\quad,
\end{equation}
\begin{equation}
L^{(x)}={\beta_x\over\epsilon_x}\left(
\begin{array}{ccc}
1 & -\gamma\phi_x & 0 \\
-\gamma\phi_x & {\gamma^2{\cal H}_x/\beta_x} & 0 \\
0 & 0 & 0
\end{array}\right)\quad,
\end{equation}
\begin{equation}
L^{(y)}={\beta_y\over\epsilon_y}\left(
\begin{array}{ccc}
0 & 0 & 0 \\
0 & {\gamma^2{\cal H}_y/\beta_y} & -\gamma\phi_y \\
0 & -\gamma\phi_y & 1
\end{array}\right)\quad.
\end{equation}
The dispersion invariant is
${\cal H}=[\eta^2+(\beta\eta^\prime-{1\over 2}\beta^\prime\eta)^2]/\beta$,
and $\phi=\eta^\prime-{1\over 2}\beta^\prime\eta/\beta$, where
$\beta$ and $\eta$ are the beta and dispersion lattice functions.

For unbunched beams $\sigma_s$ in Eq.~\ref{BM_eq} is replaced
by $C/(2\sqrt{2\pi})$, with $C$ the circumference of the machine.

\subsection*{The Bjorken-Mtingwa Solution at High Energies}

Let us first consider $1/T_p$ as given by Eq.~\ref{BM_eq}.
We first notice that, for normal storage ring lattices
(where $\langle{\cal H}_{x,y}/\beta_{x,y}\rangle\ll1$), the off-diagonal
elements in $L$, $-\gamma\phi$, are small and can be set to zero.
Then all matrices are diagonal.
Let us also limit consideration to high energies,
{\it i.e.} let us assume $a$,$b\ll1$, with
\begin{equation}
a={\sigma_H\over\gamma}\sqrt{\beta_x\over\epsilon_x}\quad,\quad\quad
b={\sigma_H\over\gamma}\sqrt{\beta_y\over\epsilon_y}\quad,
\label{ab_rev2_eq}
\end{equation}
with
\begin{equation}
{1\over\sigma_H^2}= {1\over\sigma_p^2} +
{{\cal H}_x\over\epsilon_x} +
{{\cal H}_y\over\epsilon_y}\quad.
\end{equation}
Note that if $a$,$b\ll1$,
then the beam is cooler longitudinally than transversely.
If we consider, for example, KEK's ATF, a 1.4~GeV, low emittance electron
damping ring,
$\epsilon_y/\epsilon_x\sim0.01$, $a\sim0.01$, $b\sim0.1$\cite{ATF:02b}.

\begin{figure}[htb]
\centering
\includegraphics*[width=70mm]{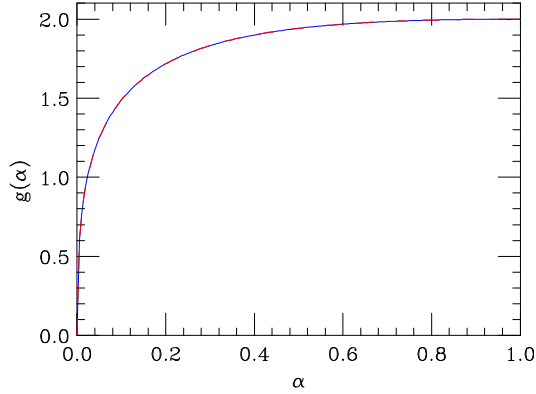}
\caption{
The auxiliary function $g(\alpha)$
(solid curve) and
an analytical approximation, $g=2\alpha^{(0.021-0.044\ln\alpha)}$ (dashes).
 }\label{hfun_fi}
\end{figure}

If the high energy conditions are met
then the 2nd term in the braces of
Eq.~\ref{BM_eq} is small compared to the first term, and can be dropped.
Now note that $L_{2,2}$ can be written as $\gamma^2/\sigma_H^2$.
For high energy beams a factor in the denominator of the integrand
of Eq.~\ref{BM_eq},
$\sqrt{\gamma^2/\sigma_H^2+\lambda}$, can be approximated
by $\gamma/\sigma_H$; also,
the (2,2) contribution to $Tr[(L+\lambda I)^{-1}]$ becomes
small, and can be set to 0.
Finally, the first of Eqs.~\ref{BM_eq} becomes
\begin{equation}
{1\over T_p}\approx  {r_0^2 cN({\rm log})\over
32\gamma^3\epsilon_x^{3/4}\epsilon_y^{3/4}\sigma_s\sigma_p^3}
\left<\sigma_H\,
g(a/b)\,\left({\beta_x\beta_y}\right)^{-1/4}\right>
\ , \label{BM_approx2_eq}
\end{equation}
with
\begin{eqnarray}
g(\alpha)&=&{4\sqrt{\alpha}\over\pi}\int_0^\infty {dy\,y^2\over
\sqrt{(1+y^2)(\alpha^2+y^2)}}\ \times
\nonumber\\
& &\times \ \left({1\over 1+y^2} + {1\over \alpha^2+y^2}
\right)\quad.
\end{eqnarray}
A plot of $g(\alpha)$ over the interval
[$0<\alpha<1$] is given in Fig.~\ref{hfun_fi};
to obtain the results for $\alpha>1$, note that
$g(\alpha)=g(1/\alpha)$.
A fit to $g$,
\begin{equation}
g(\alpha)\approx2\alpha^{(0.021-0.044\ln\alpha)}\quad\quad
[{\rm for}\ 0.01<\alpha<1]
\quad,
\end{equation}
is given by the
dashes in Fig.~\ref{hfun_fi}.
The fit has a maximum error of 1.5\% over [$0.02\leq\alpha\leq1$].

Similarly, beginning with
the 2nd and 3rd of Eqs.~\ref{BM_eq}, we obtain
\begin{equation}
{1\over T_{x,y}} \approx {\sigma_p^2\langle{\cal H}_{x,y}
\rangle\over\epsilon_{x,y}}{1\over T_p}
\quad.\label{BM_approxxy_eq}
\end{equation}
Our approximate IBS solution is Eqs.~\ref{BM_approx2_eq},\ref{BM_approxxy_eq}.
Note that
Parzen's high energy formula is a similar,
though more approximate, result to that given here\cite{Parzen:87};
and Raubenheimer's approximation
is formulas similar, though less accurate, than Eq.~\ref{BM_approx2_eq}
and identical to Eqs.~\ref{BM_approxxy_eq}\cite{Raubenheimer:91}.

Note that the beam properties in Eqs.~\ref{BM_approx2_eq},\ref{BM_approxxy_eq},
need to be the self-consistent values.
Thus, for example, to find the steady-state growth rates in electron
machines, iteration will be required.
Note also that these equations
assume that the zero-current vertical emittance is due
mainly to vertical dispersion caused by orbit errors; if it is due mainly to
(weak) $x$-$y$ coupling we
let ${\cal H}_y=0$, drop the $1/T_y$ equation, and simply let
$\epsilon_y=\kappa\epsilon_x$, with $\kappa$ the coupling factor\cite{ATF:02b}.

\section*{COMPARISON TO THE PIWINSKI SOLUTION}

\subsection*{The Standard Piwinski Solution\cite{Piwinski:99}}

The standard Piwinski solution is
\begin{eqnarray}
{1\over T_p}&=&A\left< {\sigma_h^2\over\sigma_p^2}f(\tilde a,\tilde b,q)\right>\nonumber\\
{1\over T_x}&=&A\left< f({1\over \tilde a},{\tilde b\over \tilde a},{q\over \tilde a})
+{\eta_x^2\sigma_h^2\over\beta_x\epsilon_x}f(\tilde a,\tilde b,q)\right>\nonumber\\
{1\over T_y}&=& A\left< f({1\over \tilde b},{\tilde a\over \tilde b},{q\over \tilde b})
+{\eta_y^2\sigma_h^2\over\beta_y\epsilon_y}f(\tilde a,\tilde b,q)\right>\ .
\label{tpxy_P_eq}
\end{eqnarray}
Parameters are:
\begin{equation}
{1\over\sigma_h^2}= {1\over\sigma_p^2} +
{\eta_x^2\over\beta_x\epsilon_x} +
{\eta_y^2\over\beta_y\epsilon_y}\quad,
\end{equation}
\begin{equation}
\tilde a={\sigma_h\over\gamma}\sqrt{\beta_x\over\epsilon_x},\quad
\tilde b={\sigma_h\over\gamma}\sqrt{\beta_y\over\epsilon_y},\quad
q=\sigma_h\beta\sqrt{{2d\over r_0}}\quad,
\end{equation}
The function $f$ is given by:
\begin{eqnarray}
f(\tilde a,\tilde b,q)&=&8\pi\int_0^1du\,{1-3u^2\over PQ}\ \times\nonumber\\
&  \times &\left\{2\ln\left[{q\over2}\left({1\over P}+
{1\over Q}\right)\right] -0.577\ldots\right\}\label{fabq_eq}
\end{eqnarray}
where
\begin{equation}
P^2= \tilde a^2+(1-\tilde a^2)u^2,\quad\quad Q^2= \tilde b^2+(1-\tilde b^2)u^2\ .
\end{equation}
The parameter $d$ functions as a maximum impact parameter,
 and is normally taken as the vertical beam
size.

\subsection*{Comparison of Modified Piwinski to the B-M Solution at
High Energies}

To compare with the B-M solution, let us consider a slightly
changed version of Piwinski that we call the {\it modified} Piwinski solution.
It is the standard version of Piwinski, but with $\eta^2/\beta$
replaced by ${\cal H}$ ({\it i.e.} $\tilde a$, $\tilde b$, $\sigma_h$,
become $a$, $b$, $\sigma_H$, respectively).
Let us also assume high energy beams, {\it i.e.} let $a$,$b\ll1$.

Let us sketch the derivation.
First, notice that in the integral of the auxiliary function
$f$ (Eq.~\ref{fabq_eq}):
the $-0.577$ can be replaced by 0; the $-3u^2$ in
the numerator can be set to 0;
$P$ ($Q$) can be replaced by $\sqrt{a^2+u^2}$ ($\sqrt{b^2+u^2}$).
The first term in the braces can be approximated by a constant and
then be pulled out of the integral; it becomes the effective Coulomb
log factor.
Note that for the proper choice of the Piwinski parameter $d$,
the effective Coulomb log can be made the same as the B-M parameter $({\rm log})$.
For flat beams ($a\ll b$), the Coulomb log of Piwinski becomes
$({\rm log})=
\ln{[d\sigma_H^2/(4r_0a^2)]}$.

We finally obtain
\begin{equation}
f(a,b)\approx8\pi({\rm log})\int_0^1 {du\over\sqrt{a^2+u^2}\sqrt{b^2+u^2}}\quad.
\end{equation}
The integral is an elliptic integral.
The first of Eqs.~\ref{tpxy_P_eq} then becomes
\begin{equation}
{1\over T_p}\approx  {r_0^2 cN({\rm log})\over
32\gamma^3\epsilon_x^{3/4}\epsilon_y^{3/4}\sigma_s\sigma_p^3}
\left<\sigma_H\,
h(a,b)\,\left({\beta_x\beta_y}\right)^{-1/4}\right>
\ ,\label{Tpapprox_eq}
\end{equation}
with
\begin{equation}
h(a,b)={4\sqrt{ab}\over\pi}\int_0^1
{du\over\sqrt{a^2+u^2}\sqrt{b^2+u^2}}\quad.
\end{equation}
We see that the the approximate equation for $1/T_p$ for
high energy beams according to modified Piwinski is the
same as that for B-M, except that $h(a,b)$ replaces $g(a/b)$.

\begin{figure}[htb]
\centering
\includegraphics*[width=70mm]{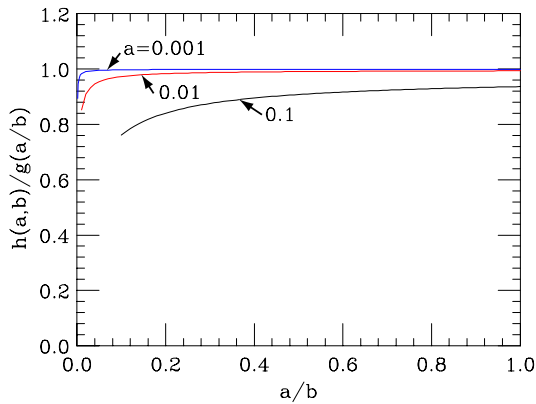}
\caption{
The ratio $h(a,b)/g(a/b)$ as function of $a/b$,
for three values of $a$.
 }\label{gfun_fi}
\end{figure}

We can now show that, for high energy beams, $h(a,b)\approx g(a/b)$:
Consider the function $\tilde h(a,b,\zeta)$, which is the same
as $h(a,b)$ except that the upper limit of integration is infinity,
and the $u^2$ in the denominator are replaced by $\zeta u^2$.
It is simple to show that
$\partial_\zeta\tilde h(a,b,\zeta)|_{\zeta=1}=g(a/b)=\tilde h(a,b,1)$.
Now for high energies ($a$,$b$ small), reducing the
upper limit in the integral of $\tilde h(a,b,1)$ to 1 does not
significantly change the
result, and $h(a,b)\approx g(a/b)$.
To  demonstrate this, we plot
in Fig.~\ref{gfun_fi} the ratio $h(a,b)/g(a/b)$ for
several values of $a$. We see, for example, for the
ATF with $\epsilon_y/\epsilon_x\sim0.01$, $a\sim0.01$, $a/b\sim0.1$,
and therefore $h(a,b)/g(a/b)=0.97$; the agreement is quite good.

Finally, for the relation between the transverse to longitudinal growth
rates according to modified Piwinski:
note that for non-zero vertical dispersion the second term
in the brackets of Eqs.~\ref{tpxy_P_eq}
(but with $\eta^2_{x,y}/\beta_{x,y}$
replaced by ${\cal H}_{x,y}$), will tend to dominate over
the first term, and the
results become the same as
for the B-M method.

In summary, we have shown that for high energy beams
($a$,$b\ll1$),
in rings with a standard type of storage ring lattice:
if the parameter $d$ in P is chosen to give the
same equivalent
Coulomb log as in B-M,
then the {\it modified} Piwinski solution agrees
with the Bjorken-Mtingwa solution.


\section*{NUMERICAL COMPARISON\cite{ATF:02b}}

We consider a numerical comparison between results of the general
B-M method, the modified Piwinski method,
and Eqs.~\ref{BM_approx2_eq},\ref{BM_approxxy_eq}.
The example is the ATF ring with no coupling and vertical dispersion
due to random orbit errors. For our example
$\langle{\cal H}_y\rangle=17$~$\mu$m, yielding a zero-current emittance
ratio of 0.7\%; the beam current is 3.1~mA. The steady-state growth rates
according to the 3 methods are given in Table~I. We note that the
Piwinski results are 4.5\% low, and the results of
Eqs.~\ref{BM_approx2_eq},\ref{BM_approxxy_eq},
agree very well with those of B-M. Finally note that, not only the
growth rates, but even the {\it differential} growth
rates---{\it i.e.} the growth rates as function of position
along the ring---agree well
for the three cases.

\begin{table}[htb]
\caption{
Steady-state IBS growth rates for an ATF example including vertical dispersion
due to random errors.}
\begin{ruledtabular}
\begin{tabular}{l c c c}
Method & $1/T_p$ [s$^{-1}$] & $1/T_x$ [s$^{-1}$] & $1/T_y$ [s$^{-1}$]\\ \hline\hline
Modified Piwinski & 25.9 & 24.7  & 18.5 \\
Bjorken-Mtingwa & 27.0 & 26.0 & 19.4\\
Eqs.~\ref{BM_approx2_eq},\ref{BM_approxxy_eq} & 27.4 & 26.0 & 19.4\\
\end{tabular}
\end{ruledtabular}
\end{table}

\begin{acknowledgments}
The author thanks K. Kubo
and A.~Piwinski for help in understanding
IBS theory.
\end{acknowledgments}

\bibliography{kbane}

\end{document}